\documentclass[aps,prb,twocolumn,showpacs,amsmath,amssymb]{revtex4}

\usepackage{graphicx}
\usepackage{dcolumn}
\usepackage{bm}
\begin{document}

\title{Vortex Dynamics at the Initial Stage of Resistive Transition
\\ in Superconductors with Fractal Cluster Structure}

\author{Yuriy I. Kuzmin}
\email{yurk@mail.ioffe.ru}

\affiliation{Ioffe Physical Technical Institute of the Russian
Academy of Sciences, 26 Polytechnicheskaya Street, Saint
Petersburg 194021 Russia}

\date{\today}

\begin{abstract}
The effect of fractal normal-phase clusters on vortex dynamics in
a percolative superconductor is considered. The superconductor
contains percolative superconducting cluster carrying a transport
current and clusters of a normal phase, acting as pinning centers.
A prototype of such a structure is YBCO film, containing clusters
of columnar defects, as well as the BSCCO/Ag sheathed tape, which
is of practical interest for wire fabrication. Transition of the
superconductor into a resistive state corresponds to the
percolation transition from a pinned vortex state to a resistive
state when the vortices are free to move. The dependencies of the
free vortex density on the fractal dimension of the cluster
boundary as well as the resistance on the transport current are
obtained. It is revealed that a mixed state of the vortex glass
type is realized in the superconducting system involved. The
current-voltage characteristics of superconductors containing
fractal clusters are obtained and their features are studied.
\end{abstract}


\maketitle

\section{Introduction}

Superconductors containing fractal clusters of a normal phase have
specific magnetic and transport properties. \cite{pla,prb} The
study of their current-voltage ({\it U-I}) characteristics
permits to get new information on the electromagnetic properties
as well as on the nature of a vortex state in such materials. The
neighborhood of resistive transition, especially its initial stage
where the energy dissipation sets-in, is of special interest. In
this region the process of vortex depinning gradually accrues
resulting finally in the destruction of superconducting state.
High temperature superconductors (HTS's) containing clusters of
correlated defects \cite{bazil,mezzetti} are of special interest
in this field. The case of clusters with fractal boundaries
provides new possibilities for increasing the critical current
value. \cite{pla2,tpl}

Let us consider the superconductor containing inclusions of a
normal phase, which are out of contact with one another. We will
suppose that the characteristic sizes of these inclusions far
exceed both the superconducting coherence length and the
penetration depth. A prototype of such a structure is a
superconducting wire.

The first generation HTS wires are fabricated following the
powder-in-tube technique (PIT). The metal tube is being filled
with HTS powder, then the thermal and deformation treatment is
being carried out. The resulting product is the wire consisting of
one or more superconducting cores sheathed by a normal metal. The
sheath endows the wire with the necessary mechanical (flexibility,
folding strength) and electrical (the possibility to release an
excessive power when the superconductivity will be suddenly lost)
properties. At present, the best results are obtained for the
silver-sheathed bismuth-based composites, which are of practical
interest for energy transport and storage. In view of the PIT
peculiarity the first generation HTS wire has a highly
inhomogeneous structure. \cite{pashitski} Superconducting core
represents a dense conglomeration of BSCCO micro-crystallites
containing normal-phase inclusions inside. These inclusions
primarily consist of a normal metal (silver) as well as the
fragments of different chemical composition, grain boundaries,
micro-cracks, and the domains of the reduced superconducting order
parameter. The volume content of a normal phase in the core is far
below the percolation threshold, so there is a percolative
superconducting cluster that carries the transport current.

The second generation HTS wires (coated conductors) have
multi-layered film structure consisting of the metal substrate
(nickel-tungsten alloy), the buffer oxide sub-layer, HTS layer
(YBCO), and the protective cladding made from the noble metal
(silver). Superconducting layer, which carries the transport
current, has the texture preset by the oxide sub-layer. In the
superconducting layer there are clusters of columnar defects that
can be created during the film growth process as well as by the
heavy ion bombardment. Such defects are similar in topology to the
vortices; therefore they suppress effectively the flux creep that
makes possible to get the critical current up to the depairing
value. \cite{mezzetti,indenbom,tonomura}

The paper is organized as follows. The setting of the problem is
described as well as vortex dynamics in percolative
superconductors in the presence of the fractal clusters is
considered in Sec. 2. The current-voltage characteristics of
superconductor with fractal cluster structure are obtained and the
peculiarities of the resistive transition in such materials are
discussed in Sec. 3.

\section{Magnetic Flux Trapping and Depinning in Percolative Superconductors}

A passage of electric current through a superconductor is linked
with the vortex dynamics because the vortices are subjected to the
Lorentz force created by the current. In its turn, the motion of
the magnetic flux transferred by vortices induces an electric
field that leads to the energy losses. In HTS's the vortex motion
is of special importance because of large thermal fluctuations and
small pinning energies. \cite{blatter} Here we will consider the
simplified model of one-dimensional line pinning when a vortex
filament is trapped on the set of pinning centers. \cite{blatter2}
Superconductors containing separated normal-phase clusters allow
for effective pinning, because the magnetic flux is locked in
these clusters, so vortices cannot leave them without crossing the
superconducting space. A cluster consists of sets of normal-phase
inclusions, united by the common trapped flux and surrounded by
the superconducting phase. The magnetic flux remains to be trapped
in the normal-phase clusters till the Lorentz force created by the
transport current exceeds the pinning force. The flux can be
created both by an external source (e.g., during magnetization in
the field-cooling regime) and by the transport current (in the
self-field regime). As soon as the transport current is turned on,
it is added to all the persistent supercurrents, maintaining the
constant magnetic flux. When the current is increased, the
vortices start to break away from the clusters of pinning force
weaker than the Lorentz force created by the transport current.
The currents are circulating around the normal-phase clusters
through the superconducting loops containing weak links. When the
total current through such a link exceeds the critical value, the
path becomes resistive and the sub-circuit involved will be
shunted by the superconducting paths where weak links are not
damaged yet. Magnetic field created by the re-distributed
transport current acts via the Lorentz force on the current
circulating around. As a result, the magnetic flux trapped therein
will be forced out through the resistive weak link, which has
become permeable to the vortices.

During this process the vortices will first pass through the weak
links, connecting the normal-phase clusters.  In this case
depinning has percolative character, \cite{yamafuji,ziese,ziese2}
because unpinned vortices move through randomly generated channels
created by weak links.  Weak links form readily in HTS's due to
the intrinsically short coherence length. \cite{sonier} Depending
on the specific weak link configuration each normal-phase cluster
has its own depinning current, which contributes to the total
statistical distribution of critical currents.

One the other hand, weak links not only allow for the magnetic
flux percolation, but they also connect superconducting domains
between themselves, maintaining the electrical current
percolation. So the composite superconductor with normal-phase
clusters represents a percolation system, where both the electric
percolation of the supercurrent and the percolation of a magnetic
flux may happen. As the transport current is increased, the local
currents flowing through ones or other weak links begin to exceed
the critical values, therefore some part of them become resistive.
Thus, the number of weak links involved in the superconducting
cluster is randomly reduced so the transition of a superconductor
into a resistive state corresponds to breaking of the percolation
through a superconducting cluster. The transport current acts as a
random generator that changes the relative fractions of conducting
components in classical percolative medium, \cite{stauffer} hence
the resistive transition can be treated as a current-induced
critical phenomenon.

Depinning current of each cluster is related to the cluster size.
Larger cluster has more weak links over its boundary, and,
consequently, the smaller depinning current. Let us take the area
of cluster cross-section as a measure of its size. This value will
be called ''the cluster area'' in the subsequent text. An
important feature of normal-phase clusters is that their
boundaries can be fractal, i. e. the perimeter of their
cross-section and the enclosed area obey the scaling law:
$P^{1/D}\propto A^{1/2}$, where $D$ is the fractal dimension of
the cluster boundary, \cite{mandelbrot} which can be fractional.
As it was first found in Ref. \cite{pla}, the normal-phase
clusters existing in superconductors can have fractal boundaries,
which has significant effect on vortex dynamics. In cited work the
fractal dimension was found as a result of geometric probability
analysis of the normal-phase clusters contained in HTS films. For
this purpose the electron photomicrographs of YBCO films prepared
by magnetron sputtering were studied. The normal-phase clusters
had the form of columnar inclusions, oriented normally to the film
surface. The profiles of cluster sections by the film plane were
clearly visible on the photomicrographs, and their geometric
probability properties were analyzed. So, instead of self-affine
fractals, \cite{mandelbrot} which the normal phase clusters are,
their cross-sections were investigated. The cluster geometric
sizes were measured by covering the digitized pictures with a
square grid. The found perimeter-area relation exhibited the
scaling behavior, which is inherent to fractals, in the range of
almost three orders of magnitude in cluster area. The fractal
dimension of the cluster boundary was estimated by a slope of the
perimeter-area regression line. The perimeter-area scaling is of
primary importance here, because porous, random, or highly
ramified clusters do not necessarily all are fractals. A fractal
cluster has such a property that its characteristic measures (in
our case - the perimeter and the enclosed area) have to obey the
certain scaling law that includes an exponent named fractal
dimension. The scaling perimeter-area behavior means that there is
no characteristic length scale over all the range of cluster
geometric sizes.

After the start of the vortex motion superconductor switches into
a resistive state. In the most practically important case of
exponential distribution of the cluster areas, which is realized
in the YBCO based film structures, \cite {pla,prb} the
distribution of critical currents is exponential-hyperbolic:
\cite{pla3}
\begin{equation}
f\left( i\right) =\frac{2C}{D}i^{-2/D-1}\exp \left(
-C\,i^{-2/D}\right)   \label{cur1}
\end{equation}
where $i\equiv I/I_{c}$ is the dimensionless electrical current
normalized to the critical current $I_{c}=\alpha \left(
C\overline{A}\right) ^{-D/2}$ of the transition into a resistive
state, $\alpha $\ is the form factor of the cluster, $C\equiv
\left( \left( 2+D\right) /2\right) ^{2/D+1}$ is the constant
depending on the fractal dimension, $\overline{A}$ is the average
cluster area, and $D$ is the fractal dimension of the cluster
boundary.

Let us note that the probability density for the critical current
distribution of Eq.~(\ref{cur1}) is equal to zero at $i=0$, which
implies the absence of any contribution from negative and zero
currents. This allows to avoid any artificial assumption about the
presence of a vortex liquid, having finite resistance in the
absence of transport currents due to presence of free vortices:
$r\left( i\rightarrow 0\right) \neq 0$. Such an assumption is
made, for example, in the case of normal distribution of critical
currents. \cite{brown}

The fractal dimension $D$ sets the perimeter-area scaling relation
that is consistent with the generalized Euclid theorem, stating
that the ratios of corresponding geometric measures are equal when
reduced to the same dimension. \cite{mandelbrot} The fractal
dimension of Euclidean clusters coincides with the topological
dimension of a line ($D=1$), while the dimension of fractal
clusters always exceeds their topological dimension ($D>1$) to
reach maximum ($D=2$) for the clusters of the most fractality. The
fractional value of dimension reflects a relationship between
characteristic measures (in what follows, the perimeter and the
enclosed area) of the object with highly indented boundary.

\section{Current-Voltage Characteristics of Superconductors with
Fractal Clusters of a Normal Phase}

The voltage drop across a superconductor in the resistive state is
an integral response of all clusters to the transport current:

\begin{equation}
u=r_{f}\int\limits_{0}^{i}\left( i-i^{\prime }\right) f\left(
i^{\prime }\right) di^{\prime }  \label{conv2}
\end{equation}
where $u$ is the dimensionless voltage and $r_{f}$ is the
dimensionless flux flow resistance. The dimensional voltage $U$
and flux flow resistance $R_{f}$ are related to the corresponding
dimensionless quantities $u$ and $r_{f}$\ by the relationship:
$U/R_{f}=I_{c}(u/r_{f})$.

Using the convolution integral of Eq.~(\ref{conv2}), we can find
the {\it U-I} characteristics of a superconductor containing
fractal clusters of a normal phase \cite{jltp}

\begin{eqnarray}
u &=&r_{f}\Biggl( i\exp \left( -C\,i^{-2/D}\right)   \nonumber \\
&&-C^{D/2}\Gamma \left( 1-\frac{D}{2},C\,i^{-2/D}\right) \Biggr)
\label{volt3}
\end{eqnarray}

\begin{figure}
\includegraphics{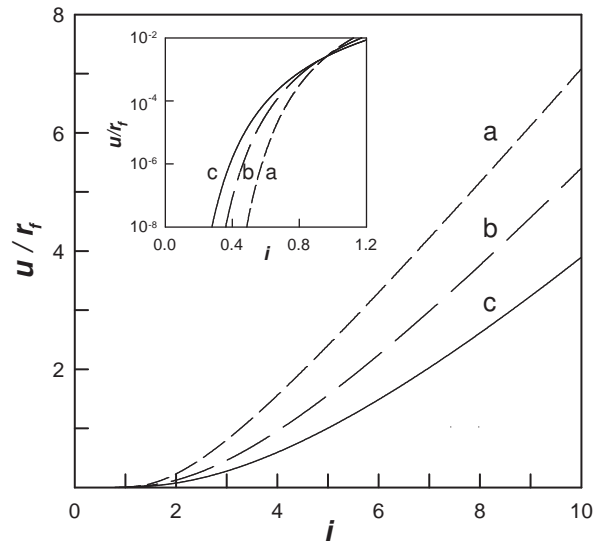}
\caption{\label{fig1} Current-voltage characteristics of
superconductor containing fractal clusters of a normal phase for
different values of a fractal dimension: (a) - $D=1$, (b) -
$D=1.5$, (c) - $D=2$. The inset shows the initial part of the
chart on a semi-logarithmic scale.}
\end{figure}

The typical curves are shown in Fig.~\ref{fig1}. This figure
demonstrates that in the range of the currents $i>1$ the
fractality of the clusters reduces the voltage arising from the
motion of the magnetic flux transferred by vortices. Meanwhile,
the situation is quite different in the neighborhood of the
resistive transition below the critical current. When $i<1$, the
higher the fractal dimension of the normal-phase cluster is, the
larger is the voltage across a sample and the more stretched is
the region of initial dissipation in {\it U-I}
characteristic. The situation is illustrated in the inset of
Fig.~\ref{fig1} where this region is shown on a semi-logarithmic
scale.

The significant difference in {\it U-I} characteristic
behavior below and above the resistive transition is related to
the dependence of free vortex density on the fractal dimension for
various transport currents. The resistive characteristics provide
additional information about the nature of the vortex state at
this stage of resistive transition. The standard parameters in
this case are {\it dc} (static) resistance $r\equiv u/i$, and {\it
ac} (differential) resistance $r_{d}\equiv du/di$. The
corresponding
dimensional quantities $R$ and $R_{d}$ can be found using the formulas $%
R=rR_{f}/r_{f}$ and $R_{d}=r_{d}R_{f}/r_{f}$, where $R_{f}$ is the
dimensional flux flow resistance.

For exponential-hyperbolic critical current distribution of
Eq.~(\ref{cur1}) the resistance of a superconductor with fractal
normal-phase clusters is given by the formulas:

\begin{eqnarray}
r &=&r_{f}\Biggl( \exp \left( -C\,i^{-2/D}\right)   \nonumber \\
&&-\frac{C^{D/2}}{i}\Gamma \left( 1-\frac{D}{2},C\,i^{-2/D}\right)
\Biggr) \label{dc4}
\end{eqnarray}

\begin{equation}
r_{d}=r_{f}\exp \left( -C\,i^{-2/D}\right)  \label{ac5}
\end{equation}
where $\Gamma (\nu ,z)$ is the complementary incomplete
gamma-function. In the limiting cases of the Euclidean clusters
($D=1$) and the clusters of the most fractal boundaries ($D=2$),
the above formulas can be simplified:
\medskip

(a) Euclidean clusters ($D=1$):

\[
r=r_{f}\left( \exp \left( -\frac{3.375}{i^{2}}\right) -\frac{\sqrt{3.375\pi }%
}{i}%
\mathrm{erfc}\left( \frac{\sqrt{3.375}}{i}\right) \right)
\]
\[
r_{d}=r_{f}\exp \left( -\frac{3.375}{i^{2}}\right)
\]
where erfc($z$) is the complementary error function, and \medskip

(b) Clusters of the most fractal boundaries ($D=2$):

\[
r=r_{f}\left( \exp \left( -\frac{4}{i}\right) +\frac{4}{i}%
\mathop{\rm Ei}%
\left( -\frac{4}{i}\right) \right)
\]
\[
r_{d}=r_{f}\exp \left( -\frac{4}{i}\right)
\]
where Ei($z$) is the exponential integral function.

\begin{figure}
\includegraphics{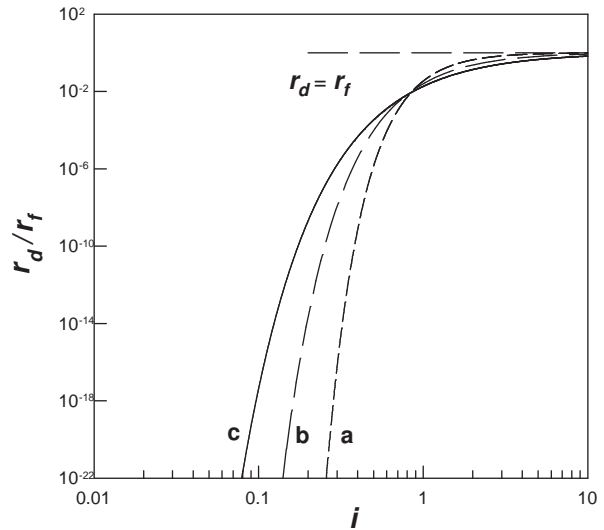}
\caption{\label{fig2} Dependence of the differential resistance of
superconductor with fractal clusters of a normal phase on a
transport current for different values of a fractal dimension: (a)
- $D=1$, (b) - $D=1.5$, (c) - $D=2$. The dashed horizontal line at
the upper right $r_{d}=r_{f}$ corresponds to the flux flow
regime.}
\end{figure}

Since the {\it U-I} characteristic of Eq.~(\ref{volt3}) is
nonlinear, the {\it dc} resistance of Eq.~(\ref{dc4}) is not
constant and depends on the transport current. The more convenient
parameter is the differential
resistance, a small-signal parameter that gives the slope of the {\it U}-%
{\it I} characteristic. Figure \ref{fig2} shows the graphs of the
differential resistance as a function of transport current for
superconductor with fractal normal-phase clusters. The curves
drawn for the Euclidean clusters ($D=1$) and for the clusters of
the most fractal boundaries ($D=2$) bound a region containing all
the resistive characteristics for an arbitrary fractal dimension.
As an example, the dashed curve shows the case of the fractal
dimension $D=1.5$. The dependencies of resistance on the current
shown in Fig.~\ref{fig2} are typical of the vortex glass, inasmuch
as the curves plotted on a double logarithmic scale are convex and
the resistance tends to zero as the transport current decreases,
$r_{d}\left( i\rightarrow 0\right) \rightarrow 0 $, which is
related to the flux creep suppression. \cite{brown,blatter} A
vortex glass represents an ordered system of vortices without any
long-range ordering. At the same time, the vortex configuration is
stable in time and can be characterized by the order parameter of
the glassy state. \cite{fisher,fisher2} In the {\it H}-{\it T}
phase diagram, mixed state of the vortex glass type exists in the
region below the
irreversibility line. The dashed horizontal line at the upper right of Fig.~%
\ref{fig2} corresponds to a viscous flux flow regime ($r_{d}=r_{f}=const$%
), which can only be approached asymptotically.

The resistance of superconductor is determined by the density $n$
of free vortices broken away from pinning centers by the transport
current $i$

\begin{equation}
n\left( i\right) =\frac{B}{\Phi _{0}}\int\limits_{0}^{i}f\left(
i^{\prime }\right) di^{\prime }=\frac{B}{\Phi _{0}}\exp \left(
-C\,i^{-2/D}\right) \label{dens6}
\end{equation}
where $B$ is the magnetic field, $\Phi _{0}\equiv hc/\left(
2e\right) $ is the magnetic flux quantum, $h$ is the Planck
constant, $c$ is the speed of light, and $e$ is the electron
charge. The more vortices are free to move, the stronger is the
induced electric field, and therefore, the higher is the voltage
across a sample at the same transport current. A comparison of
expressions of Eqs.~(\ref{ac5}) and (\ref{dens6}) shows that the
differential resistance is proportional to the density of free
vortices: $r_{d}=(r_{f}\Phi _{0}/B)n$. Resistance of the
superconductor in a resistive state is determined by the motion
just of these vortices.

\begin{figure}
\includegraphics{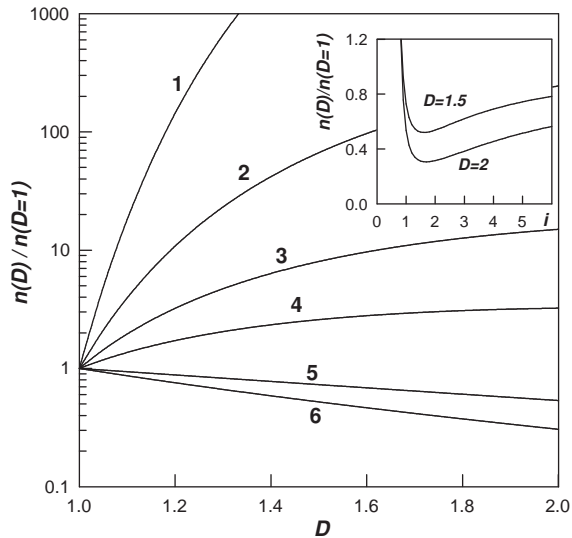}
\caption{\label{fig3} Dependence of the free vortex density on the
fractal dimension for different values of a transport current: (1)
- $i=0.4$, (2) - $i=0.5$, (3) - $i=0.6$, (4) - $i=0.7$, (5) -
$i=1$, (6) - $i=1.6875$. The inset shows the free vortex density
versus current for two different values of a fractal dimension.}
\end{figure}

Figure \ref{fig3} demonstrates dependence of the relative density
of free vortices $n(D)/n(D=1)$ (relatively to the value for
clusters with Euclidean boundary) on the fractal dimension for
different values of transport currents. The vortices are broken
away from pinning centers mostly when $i>1$, that is to say, above
the resistive transition. Here the free vortex density decreases
with increasing the fractal dimension. Such a behavior can be
explained by the fact that the critical current distribution of
Eq.~(\ref{cur1}) broadens out, moving towards greater magnitudes
of current as the fractal dimension increases. It means that more
and more clusters of high depinning current, which can trap the
vortices best, are being involved in the process. The smaller part
of the vortices is free to move, the smaller is the induced
electric field. An important feature of the fractal
superconducting structures is that fractality of cluster boundary
enhances pinning \cite{tpl} and, hence, a current-carrying
capability of the superconductor. This is demonstrated both in
Fig.~\ref {fig2}, where resistance decreases with increasing
fractal dimension above the resistive transition. The relative
change in free vortex density depends on the transport current
(see inset in Fig.~\ref{fig3}) and in the limiting case of the
most fractal boundary $D=2$ reaches a minimum for $i=1.6875$
(curve 6 in Fig.~\ref{fig3} goes below others). That corresponds
to the maximum pinning gain and the minimum level of dissipation.
As may be seen in Fig.~\ref{fig1}, the voltage across a sample
carrying the same transport current decreases with increasing
fractal dimension.

In the range of transport currents below the resistive transition
($i<1$), the situation is different: resistance, as well as the
free vortex density, increase for the clusters of greater fractal
dimension (see Figs.~\ref{fig2} and \ref{fig3}). Such a behavior
is related to the fact that the critical current distribution of
Eq.~(\ref{cur1}) broadens out, covering both high and small
currents as the fractal dimension increases. For this reason, the
breaking of the vortices away under the action of transport
current begins earlier for the clusters of greater fractal
dimension. In spite of sharp increase in relative density of free
vortices (Fig.~\ref{fig3}), the absolute value of vortex density
in the range of currents involved is very small (much smaller than
above the resistive transition).  So the vortex motion does not
lead to the destruction of superconducting state yet, and the
resistance remains very low.  The low density of vortices at small
currents is related to the peculiarity of exponential-hyperbolic
distribution of Eq.~(\ref{cur1}). This function is so ``flat'' in
the vicinity of the coordinate origin that all its derivatives are
equal to zero at the point of $i=0$: $d^{k}f(0)/di^{k}=0$ for any
value of $k$.  This mathematical feature has a clear physical
meaning: so small a transport current does not significantly
affect the trapped magnetic flux because there are scarcely any
pinning centers of such small critical currents in the overall
statistical distribution, so that nearly all the vortices are
still pinned. This interval corresponds to the so-called initial
fractal dissipation regime, which was observed in BPSCCO samples
with silver inclusions as well as in polycrystalline YBCO and
GdBCO samples.  \cite{prester} As for any hard superconductor some
dissipation in a resistive state does not mean the destruction of
phase coherence yet. Dissipation always accompanies any motion of
vortices that can happen in hard superconductor even at low
transport current. Superconducting state collapses only when a
growth of dissipation becomes avalanche-like as a result of
thermo-magnetic instability.

\section{Conclusion}

The fractal properties of the normal-phase clusters significantly
affect the resistive transition. This phenomenon is related to the
features of the critical current distribution. The current-voltage
and resistance characteristics of superconductor with fractal
cluster structure correspond to a mixed state of the vortex glass
type. An important result is that the presence of the fractal
clusters of a normal phase in a superconductor enhances the
pinning. This feature provides the principally new possibility for
increasing the current-carrying capability of a superconductor
without changing of its chemical composition.

\begin{acknowledgments}
This work is supported by the Saint Petersburg Scientific Center
of the Russian Academy of Sciences.
\end{acknowledgments}

\bibliography{ARW2}

\end{document}